\documentstyle[12pt,twoside,fleqn,epsfig,espcrc1]{article}

\def\Journal#1#2#3#4{{#1} {\bf #2}, #3 (#4)}

\def\PRC{{Phys. Rev.} C}

\def\a{\alpha}
\def\e{\epsilon}
\def\q{\theta}
\def\t{\tau}

\def\be{\begin{equation}}
\def\ee{\end{equation}}
\def\bea{\begin{eqnarray}}
\def\eea{\end{eqnarray}}
\def\eref#1{Eq.~(\ref{#1})}

\def\fref#1{Fig.~\ref{#1}}
\def\bfig{\begin{figure}}
\def\efig{\end{figure}}

\newcommand{\lan}{\langle}
\newcommand{\ran}{\rangle}

\title{Equilibration and Particle Production in an Increasingly
       Strongly Interacting Parton Plasma} 

\author{S.M.H. Wong\address{Fachbereich Physik, Universit\"at Wuppertal,
                            D-42097 Wuppertal, Germany}%
        \thanks{The author is grateful for local financial support from the 
                Yamada Science Foundation during the attendance of this 
                conference.}}

\begin{document}


\maketitle

\begin{abstract}
We report on a new equilibration scenario in relativistic heavy
ion collisions, the scenario of the Increasingly Strongly 
Interacting Parton Plasma, and the effects of this scenario on
equilibration and open charm, photon and dilepton production. 
The parton plasma is shown to be a very special kind of many-body 
system, which contains new physics concerning the approach 
towards equilibrium. This is likely to be unique to the parton
plasma. \hfill{\footnotesize \sffamily WU-B 98-1}
\end{abstract}

\section{Introduction}
\label{sec:intro}

In this talk, we present a new equilibration scenario, which
emerges out of the initially gluon dominated hot gluon
and progressively gains importance in time, this is the
scenario of the Increasingly Strongly Interacting Parton
Plasma (ISIPP). The new scenario arises out of the previously
neglected running of the coupling. In high energy nuclear 
collisions, the interactions in the early stage are
pretty hard, so perturbative QCD is applicable. It is
common therefore to choose a small {\em fixed} coupling
of $\a_s=0.3$, with $\Lambda_{QCD}=$200 MeV, this corresponds 
to a {\em fixed} average momentum transfer of 2 GeV.
Now for a very typical interaction, a simple t-channel
exchange for parton 1 and 2 with 4-momenta $p_1$ and $p_2$
and final parton 3 and 4 with 4-momenta $p_3$ and $p_4$, 
respectively, the momentum transfer $Q^2$ would be bounded
between $0 < Q^2 < 4 p_1 p_3$. For very typical partons, the 
upper limit would be given by $4 p_1 p_3 = 4 \lan \e \ran^2$,
essentially the squared of the average parton energy.
In ref. \cite{wong3}, we showed that the time evolution of
$\lan \e \ran$ in a plasma time-evolved with a fixed 
$\a_s=0.3$ at LHC and at RHIC energies. The change of 
$\lan \e \ran$ from the beginning to the end of the 
evolution is a decrease of at least 1 GeV and nearly 
2 GeV at LHC. So the average momentum transfer $Q$ in the
system will decrease in time. This leads to an increase
in the strength of the interactions and the interactions
in the parton plasma will become stronger and stronger
as the system evolves. This is the origin of the
new scenario of ISIPP. In the following, we discuss but
not show, due to lack of space, the effects on equilibration
of the parton plasma in the new scenario. We show that 
the parton plasma is a special kind of many-body system 
unlike ordinary matter and is probably unique. To end, we 
have three news to report on three different particle 
productions.

\section{Effects of ISIPP on Equilibration}
\label{sec:equil}

By combining Boltzmann equation with the relaxation time
approximation for the collision terms, 
\be C(p,\t) = - \frac{f(p,\t)-f_{eq}(p,\t)}{\q (\t)}
\label{eq:relax}
\ee
and construction of the latter again explicitly from
QCD matrix elements of the simplest elastic and
inelastic interactions, we can solve for the distribution 
function $f(p,\t)$ and hence find out the time evolution
of the parton plasma. This has been done in ref. \cite{wong2}
for a plasma with fixed interaction strength. For ISIPP, 
one must take into account also the decrease of the 
average energy of the system so this is accompanished
by using $Q = \lan \e \ran$, a not too unreasonable choice,
and inserting this into the one-loop running coupling formula
to obtain a time-dependent $\a_s$. With this approach, we 
found that both chemical and kinetic equilibration of 
quarks and gluons in the plasma are speeded up but only 
for quark and antiquark are there any improvements \cite{wong3} 
towards the end of the time evolution. We refer the readers 
to \cite{wong3} for details. But still viewed on the 
whole, the equilibration of the system is faster and 
improved. The only drawback in the time evolution
of the system is the more rapid cooling and hence 
shortened duration of the parton phase due to energy 
sharing from enhanced particle creation and longitudinal
expansion. This reduction when going from the standard plasma 
to ISIPP is rather large. This is determined by the moment 
when the estimated parton temperatures reach 
$T_c=200$ MeV, the end point of the time evolution. 
The reduction at LHC is as much as 5.0 fm/c and 3.5 fm/c
at RHIC. These are the main effects on equilibration in 
ISIPP.

\section{Parton Plasma is a Unique Many-Body System}
\label{sec:unique}
\addtolength{\textheight}{1ex}

We show in this section that the parton plasma is a special
kind of many-body system. The quantity that allows us
to reveal the secret of the parton plasma is the collision 
time $\q$ as appeared in \eref{eq:relax}. The inverse of 
this quantity measures the net interaction rate which is the
difference between the rate of the collisions going 
forward and backward. As the plasma is being driven out 
of equilibrium, this difference of the collision rate must 
increase, which is precisely what going out of equilibrium 
means. So the net reaction rate must go up and $\q$ 
must decrease when the system is going out of equilibrium.
When the system is in equilibrium, detail balance means
the difference of the collision rate is zero. So when the
system approaches equilibrium, the net rate must
go down and $\q$ must therefore increase. So for our
situation, $\q$ must first decrease, makes a turn and
increases again.
\bfig
\centerline{
\hbox{
\epsfig{file=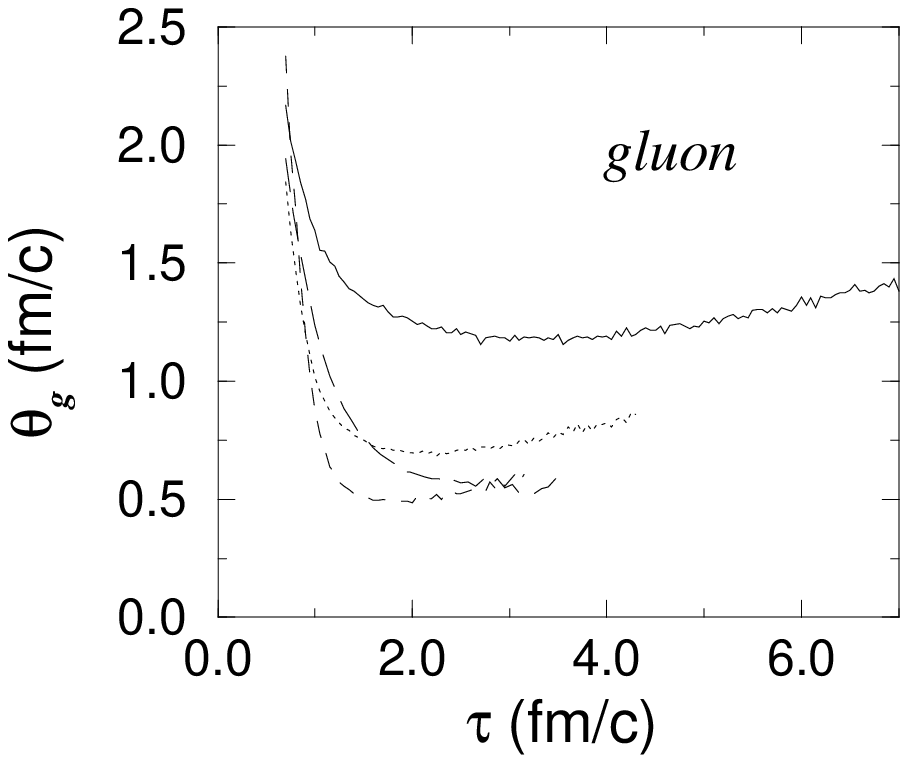,width=2.3in} \ \ \ \ \ 
\epsfig{file=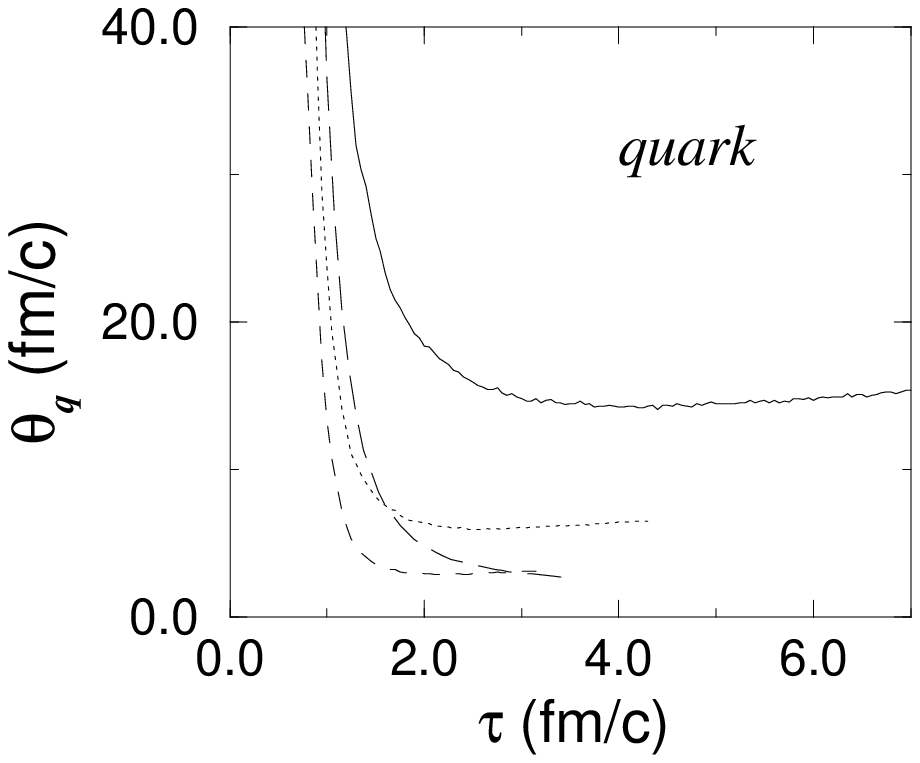,width=2.3in}
}}
\vskip -0.80cm
\caption{The time evolution of the collision times at RHIC
for gluon and quark in ISIPP (long dashed) is different
from parton plasma with fixed $\a_s=$0.3 (solid), 0.5 (dotted) 
and 0.8 (dashed). It shows the unique approach of ISIPP towards
equilibrium.}
\label{fig:coll}
\vskip -0.50cm
\efig
In \fref{fig:coll}, we show the collision times for quark 
and gluon at RHIC for various fixed couplings and for 
ISIPP. As can be seen, for the plasmas with fixed
coupling, the behaviour of $\q$ follows what we have
just described, but for ISIPP, the $\q$'s tend to 
continue to decrease with time. This decrease is clearly
of a different type from the initial drop when the plasma
is being driven out of equilibrium. Since $\q$'s for ISIPP
are not rising in the later stage in general, it seems
to mean that the plasma is not equilibrating. This clearly
contradicts what we have already said about ISIPP in 
Sec. \ref{sec:equil}. What is happening is, in fact, in the 
above difference of collision rate, a power of $\a_s$ can be 
extracted out as a prefactor, the increasing coupling can 
therefore compensate for the decrease in the difference as 
equilibrium is approached. The result is, equilibration 
proceeds faster and faster in ISIPP whereas,
in ordinary many-body system, it is slower and slower.
This is a new physics of the parton plasma that, as
far as we know, has never been revealed before, which
gives the ISIPP a unique status as a many-body system.

\section{The 3 News of Particle Production in ISIPP}
\label{sec:news}

When we go from the standard parton plasma to ISIPP,
there are two categories of effects that make the
difference in particle production. They are
the direct and indirect effects \cite{wong4}. The former 
is where QCD has at least a partial role to play in the
production process and hence there will be explicit
dependence on $\a_s$. The latters are where the
coupling effects first act on the time evolution of the
system which then get passed onto the production process.
The reduced duration of the parton phase already mentioned
and enhanced or reduced parton densities \cite{wong3} are
examples of the latter. The first is of importance for
particle production because of the need to integrate
over spacetime history of the collisions and the second
is obvious. We consider 3 types of particle production
and we only show the $p_T$ or invariant mass $M$ distributions
at LHC.

\bfig
\centerline{
\hbox{
\epsfig{file=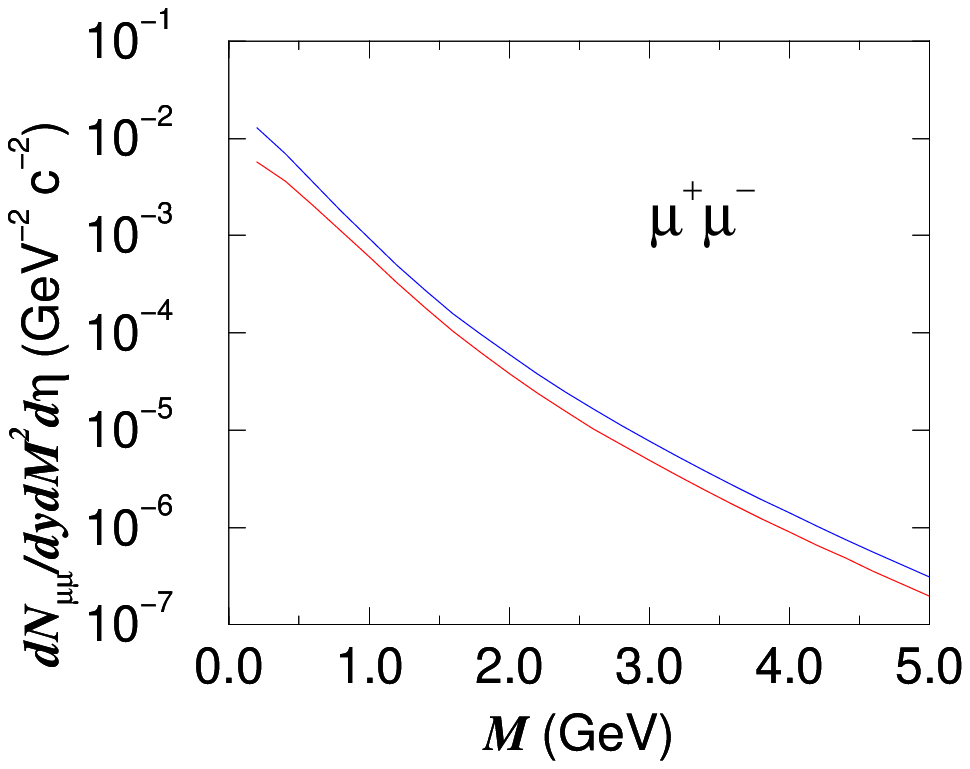,width=2.3in} \ \ \ \
\epsfig{file=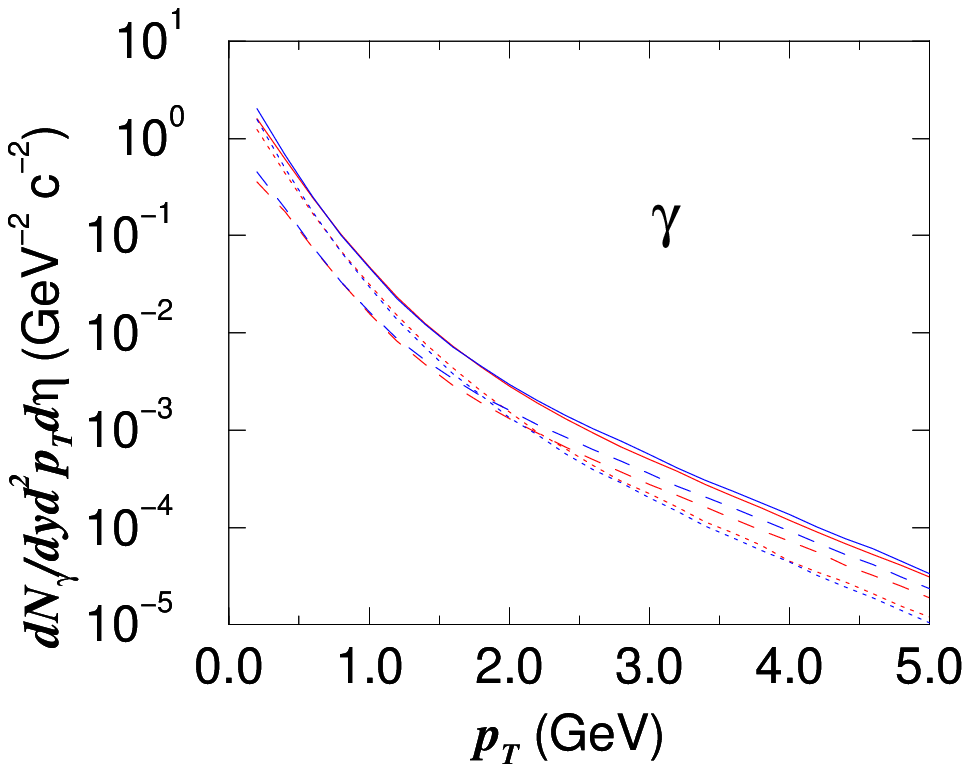,width=2.3in}
}}
\centerline{
\epsfig{file=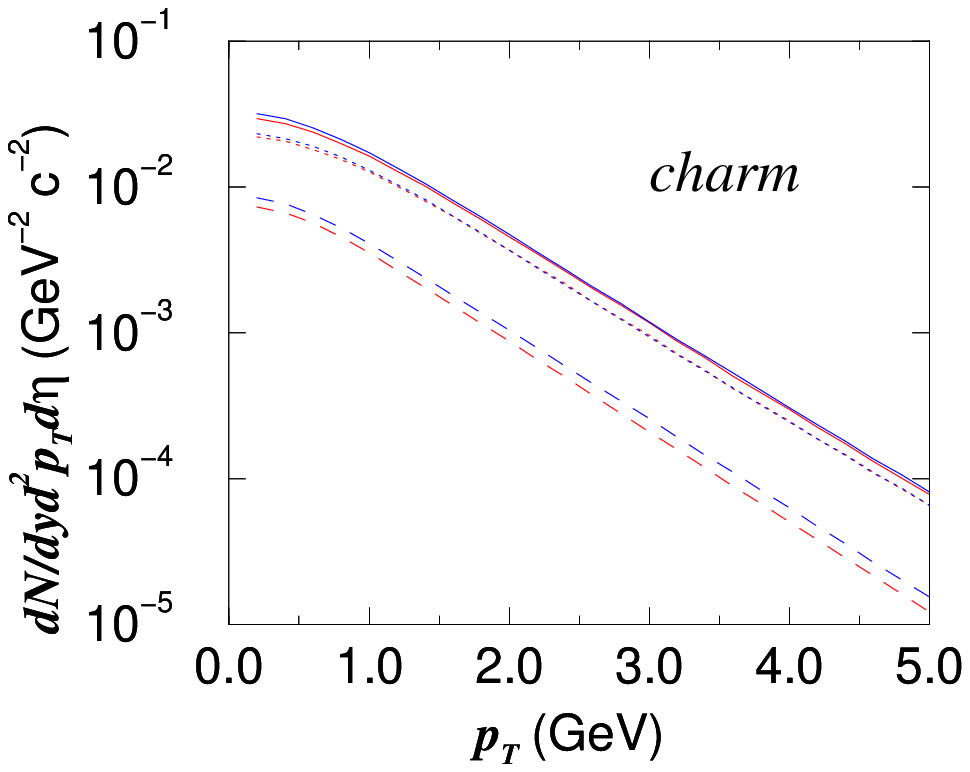,width=2.3in}
}
\vskip -1.0cm
\caption{Comparison of dilepton, photon and open charm 
production at LHC from ISIPP with the standard parton plasma.
For dilepton, the lower curve is from ISIPP. For photon, 
the total sum (solid), contribution from Compton scattering 
(dotted) and annihilation (dashed) are essentially the same 
in the two plasmas. The total (solid) and gluon conversion 
(dotted) to open charm from ISIPP are the same in the
two scenarios, while annihilation (dashed) is slightly
down in ISIPP.}
\label{fig:pp}
\vskip -0.70cm
\efig

\begin{itemize}
\setlength{\itemsep}{-0.5ex}

\item[(i)]{First News --- Dilepton Production \\
At leading order, this is only subject to indirect effects.
As seen in \fref{fig:pp}, reduced production time is more
important than enhanced fermion densities and therefore the
dilepton production in ISIPP is suppressed.}

\item[(ii)]{Second News --- Photon Production \\
Photon production comes from Compton and annihilation
contribution. This time both direct and indirect effects
are at work. Photon production plot in \fref{fig:pp} shows 
that negative and positive direct and indirect effects 
largely cancel out each other so the $p_T$ distribution 
remains essentially unchanged.}

\item[(iii)]{Third News --- Open Charm Production \\
This process is not subject to direct effect, contrary to
appearance, because this is a hard process and is therefore 
at a different scale, in general, from that for equilibration 
\cite{wong4}. As seen in \fref{fig:pp}, the total yield is 
essentially the same in ISIPP over the standard plasma. The 
ability of open charm as a probe is therefore unaffected.}
\end{itemize}

\section{Summary and Outlook}
\label{sec:sum}

We have pointed out that the ISIPP scenario is a necessary
stage in heavy ion collisions and equilibration is faster
and improved as a result but the parton phase is shortened. 
The approach towards equilibrium for ISIPP is different
from other many-body system and contains new physics. 
Particle productions at leading order in the new plasma
have mixed results. In view of the increasing coupling, 
higher orders are more important for electromagnetic
radiations. There is even a chance of enhancement. We have 
not mentioned the effect of reduction of generated entropy
in ISIPP \cite{wong3}, this will lead to a shortened
mixed phase in the case of a first order phase transition 
and therefore reduced final hadron multiplicity and 
hadron gas density and therefore reduced hadronic signals. 
So enhanced partonic but reduced hadronic signals, for example
the electromagnetic ones suggested here, that follow from ISIPP, 
if confirmed, would contribute positively towards the 
possibility of having the quark-gluon plasma outshining 
hadronic matter.

\end{document}